\documentclass[11pt]{article}

\usepackage{amsmath,amssymb,latexsym}

\def\ol{\overline}

\def\Cont{{\bf Cont}}

\newtheorem{thm}{\bf Theorem}[section]
\newtheorem{lem}[thm]{\bf Lemma}        
\newtheorem{prop}[thm]{\bf Proposition}  
\newtheorem{cor}[thm]{\bf Corollary}

\newtheorem{remark}[thm]{\bf Remark}
\newtheorem{expl}[thm]{\bf Example}

\def\Bbox{
{\unskip\nobreak\hfil\penalty50
\hskip1em\hbox{}\nobreak\hfil{\lower .5pt \hbox{$\Box$}}
\parfillskip=0pt \finalhyphendemerits=0 \par}
}

\def\eop{
\ifmmode {\hbox{\Bbox}} \else \Bbox \fi
}

\def\bbox{
\ifmmode {\hbox{\bbox}} \else \Bbox \fi
}

\def\ol#1{\overline{#1}}

\title{{\bf A connection between concurrency and language theory}}

\author{Z. \'Esik
\\
Dept. of Computer Science\\
National University of Excellence of Szeged\\
6720 Szeged\\
\'Arp\'ad t\'er 2}

\date{\empty}
\usepackage{amsmath, latexsym}

\def\os{{\oplus}}

\usepackage{amsmath, amssymb}

\def\BRST{{\bf BRST}}
\def\BST{{\bf BST}}
\def\RST{{\bf RST}}
\def\SRST{{\bf SRST}}
\def\Lang{{\bf Lang}}
\def\TreeLang{{\bf TreeLang}}
\def\Words{{\bf W}}
\def\Tree{{\bf Tree}}
\def\ST{{\bf ST}}
\def\SST{{\bf SST}}
\def\N{{\mathbb{N}}}
\def\ex{{\sf ex}}

\def\CFL{{\bf CFL}}
\def\Reg{{\bf Reg}}

\def\FST{{\bf FST}}
\def\CF{{\bf CFL}}
\def\SFST{{\bf SFST}}
\def\omegaSFST{{{\omega}{\bf SFST}}}

\input{diagxy}

\def\1#1{
\hbox{{\bf 1}$_{#1}$}
}

\begin{document}

\maketitle

\begin{abstract}
We show that three fixed point structures equipped with (sequential) composition,
a sum operation, and a fixed point operation share the same valid 
equations. These are the theories of (context-free) languages, (regular) 
tree languages, and simulation equivalence classes of (regular) synchronization 
trees (or processes). The results reveal a close relationship between
classical language theory and process algebra.
\end{abstract}

\section{Introduction}

Iteration theories \cite{BEbook} capture the equational properties of fixed point operations, including the least fixed point operation over continuous or monotone functions over cpo's or complete lattices, the initial fixed point operation over 
continuous functors over categories with directed colimits, Elgot's (pointed) 
iterative theories \cite{Elgot}, and many other structures. 
It was argued in \cite{BEbook,BEtutorial} that all natural cartesian fixed point models 
lead to iteration theories, and it was proved in \cite{PlotkinSimpson} by rigorous means
that essentially all natural cartesian fixed point models satisfy exactly the 
equations of iteration theories.

But several models have an additional structure, such as an additive structure, which interacts with the cartesian operations and the fixed point operation in a nontrivial way. 
The relationship between the iteration theory structure and the additional 
operations has been the subject of several papers, including
\cite{Acetoetal,BEregular,BET,Essup,BErational,Esregtree1,Esinj,Esregtree2,EKsemialg,EKinductive} 
and the more recent \cite{Esweighted}.  
In many cases, it was possible to capture this relationship by a finite number of equational 
(or sometimes quasi-equational)
axioms. As a byproduct of these results, it was possible to give complete sets of 
equational axioms for various bisimulation and trace based process behaviors, rational power series and regular languages, regular tree languages, and many other models.

The theory of simulation equivalence classes of (regular) synchronization trees over 
a set of action symbols, equipped with the cartesian operations, the least fixed point operation 
\emph{and sum}, has a finite equational axiomatization relatively to iteration theories \cite{Essup}. 
Incidentally, the very same equations hold for continuous or monotone functions over complete lattices 
equipped with the least fixed point operation and the pointwise binary supremum operation as sum,
or more generally,  in all `($\omega$-)continuous idempotent grove theories'.
In this paper, our main new contribution is that two more well-known 
classes of structures relevant to computer science
are of this sort, the theories of 
(regular) tree languages and the theories of (context-free) languages (Theorem~\ref{thm-main}).  
In our argument, we will make use of a concrete characterization of the free 
$\omega$-continuous idempotent grove theories, which is a result of independent interest
(cf. Theorem~\ref{thm-free-omegaideal}).
The facts proved in the paper reveal a close relationship between models of concurrency, automata and language theory, and models of denotational semantics.

The results of this paper can be formulated in several different formalism including 
`$\mu$-terms', `letrec expressions', or cartesian categories. We have chosen the 
simple language of Lawvere theories, i.e., cartesian categories generated by a single 
object. The extension of the results to many-sorted theories is straightforward.

\section{Theories}

In any category, we write the composition $f\cdot g$ of morphisms 
$f: a \rightarrow b$ and $g: b \rightarrow c$ in diagrammatic order, and we 
let $\1{a}$ denote the identity morphism $a \rightarrow a$. 
For an integer $n\geq 0$, we let $[n]$ denote the set $\{1,\ldots,n\}$.
When $n = 0$, this set is empty.

A  \emph{(Lawvere) theory} \cite{Lawvere,BEbook} is a small category $T$ whose objects 
are the nonnegative integers such that each object $n$ is the $n$-fold coproduct
of object $1$ with itself. The home-set of morphisms $n \rightarrow p$ of a theory $T$ 
is denoted $T(n,p)$. We assume that every theory comes 
with distinguished coproduct injections $i_n : 1 \rightarrow n$, 
$i \in [n]$, $n \geq 0$. Thus, for any sequence of morphisms $f_1,\ldots,f_n: 1 \rightarrow p$, 
there is a unique morphism $f : n \rightarrow p$ with $i_n \cdot f = f_i$, for all $i \in [n]$. 
We denote this unique morphism $f$ by $\langle f_1,\ldots,f_n\rangle$ and call it the 
\emph{tupling} of the $f_i$. When $n = 0$, we also write $0_p$. 
Since $0$ is initial object, $0_p$ is the unique morphism $0 \rightarrow p$.
It is clear that $\1{n} = \langle 1_n,\ldots,n_n\rangle$
for all $n\geq 0$. We require that $\1{1} = 1_1$, so that $\langle f \rangle = f$ 
for all $f: 1 \rightarrow p$. Since the object $n + m$ is the coproduct of objects $n$ 
and $m$ with respect to the coproduct injections 
\begin{eqnarray*}
\kappa_{n,n+m}  &=& \langle 1_{n+m},\ldots,n_{n+m}\rangle:  n \rightarrow n+ m\\
\lambda_{m,n+m} &=& \langle (n+1)_{n+m},\ldots,(n+m)_{n+m}\rangle:  m \rightarrow n+m,
\end{eqnarray*}
each theory is equipped with a \emph{pairing} operation 
mapping a pair of morphisms $(f,g)$ with $f: n \rightarrow p$ and 
$g: m \rightarrow p$ to $\langle f,g \rangle : n + m \rightarrow p$:
$$
\bfig
\Vtrianglepair/>`<-`>`>`>/[n`n+m`m`p;\kappa`\lambda`f`\langle f,g \rangle `g]
\efig 
$$
The pairing operation is associative and satisfies $\langle f,0_p\rangle = f = \langle 0_p,f\rangle$ for all $f: n \rightarrow p$. 
Also, we can define for $f: n \rightarrow p$ and $g: m \rightarrow q$ the 
morphism $f \oplus g: n + m \rightarrow p + q$ as 
$\langle f \cdot \kappa_{p,p+q},\,g \cdot \lambda_{q,p+q} \rangle$.
Then $f \oplus g$ is the unique morphism $n+m \rightarrow p+q$ 
with 
\begin{eqnarray*}
\kappa_{n,n+m}\cdot (f \oplus g) &=& f \cdot \kappa_{p,p+q}\\
\lambda_{m,n+m}\cdot (f \oplus g) &=& g \cdot \lambda_{q,p+q}.
\end{eqnarray*}
$$
\bfig
\square|almb|[n`n+m`p`p+q;\kappa`f`f\os g`\kappa]
\square(500,0)/<-``>`<-/[n+m`m`p+q`q;\lambda``g`\lambda]
\efig
$$
 The $\oplus$ operation is associative, and $0_0 \oplus f 
= f = f \oplus 0_0$ for all $f: n \rightarrow p$. Also, 
\begin{eqnarray*}
(f \oplus g)\cdot \langle h,k\rangle &=& \langle f\cdot h, \, g \cdot k\rangle
\end{eqnarray*}
for all $f: n \rightarrow p$, $g: m \rightarrow q$, $h: p \rightarrow r$ and $k: q \rightarrow r$.

Each theory $T$ may be seen as a many-sorted algebra,
whose set of sorts is the set 
$\N \times \N$ of all ordered pairs of nonnegative integers,
 satisfying the following 
equational axioms:
\begin{eqnarray*}
(f\cdot g) \cdot h &=& f \cdot (g \cdot h),\quad f: m \rightarrow n,\ g: n \rightarrow p,\ h: p\rightarrow q\\
\1{n}\cdot f &=& f \ =\ f \cdot \1{p},\quad f: n\rightarrow p\\
i_n \cdot \langle f_1,\ldots,f_n\rangle &=& f_i,\quad f_1,\ldots,f_n: 1 \rightarrow p,\ i\in [n],\ n\geq 0\\
\langle 1_n \cdot f,\ldots, n_n\cdot f \rangle &=& f, \quad f: n \rightarrow p\\
\1{1} &=& 1_1.
\end{eqnarray*}
Morphisms of theories are functors preserving objects and distinguished morphisms.
It follows that any theory morphism preserves the tupling,
pairing and $\oplus$ operations. The kernels of theory morphisms 
are called {\em theory congruences}. The {\em quotient} $T/\equiv$ of a theory
$T$ with respect to a theory congruence is defined as usual. A \emph{subtheory} of
a theory $T$ is a theory $T'$ whose set of morphisms is included in the 
morphisms of $T$ such that the natural embedding of $T'$ into $T$ is a theory morphism
$T' \rightarrow T$. See \cite{BEbook} for more details.

We end this section by providing some examples.

Let $X = \{x_1,x_2,\ldots\}$ denote a fixed countably infinite set
of variables, and let $A$ be a set disjoint with $X$. For each
$p \geq 0$, let $X_p = \{x_1,\ldots,x_p\}$. 
The theory $\Words_A$ has as morphisms $1 \rightarrow p$
all words in $(A \cup X_p)^*$. 
A morphism $n \rightarrow p$ is an $n$-tuple of morphisms
$1 \rightarrow p$. For morphisms $u = (u_1,\ldots,u_n): n \rightarrow p$ 
and $v = (v_1,\ldots,v_p): p \rightarrow q$, 
we define $u \cdot v = (u_1\cdot v,\ldots,u_n\cdot v)$, where for each $i\in [n]$, $u_i \cdot v$ 
is the word obtained from $u_i$ by substituting a copy of $v_j$ for each occurrence of the 
variable $x_j$ in $v_i$, for all $j \in [p]$. Equipped with this composition 
operation and the morphisms $\1{n} = (x_1,\ldots,x_n):n \rightarrow n$ as identity morphisms, 
$\Words_A$ is a category. In fact, $\Words_A$ is a theory with distinguished morphisms 
$i_n = x_i: 1 \rightarrow n$, $i \in [n]$, $n \geq 0$.

Suppose now that $\Sigma = \biguplus_{k \geq0} \Sigma_k$ is a ranked set which is disjoint with $X$. 
We may view $\Sigma$ as a pure set and form the theory $\Words_\Sigma$.
Consider the subtheory $\Tree_\Sigma$ of $\Words_\Sigma$ consisting
of the $\Sigma$-trees (or $\Sigma$-terms). A morphism $1 \rightarrow p$ 
in $\Tree_\Sigma$ is a well-formed word in $(A \cup X_p)^*$ which is either 
a variable in $X_p$ or a word of the form $\sigma t_1\ldots t_k$ for a letter $\sigma \in \Sigma_k$ 
and trees $t_1,\ldots,t_k: 1 \rightarrow p$. A morphism $n\rightarrow p$ is an $n$-tuple of morphisms $n \rightarrow p$. It is well-known that the theory $\Tree_\Sigma$ is the free theory,
freely generated by $\Sigma$. Indeed, each letter $\sigma\in \Sigma_n$ may be identified
with a tree in $\Tree_\Sigma(1,n)$ so that given any theory $T$ and rank preserving 
function $\varphi: \Sigma \rightarrow T$, there is a unique theory morphism
$\varphi^\sharp: \Tree_\Sigma \rightarrow T$ extending $\varphi$.

If in the previous example $\Sigma$ is empty, then we obtain the initial 
theory $\Theta$. A morphism $n \rightarrow p$ of this initial theory 
is a tupling of distinguished morphisms and may be identified 
with a function $[n]\rightarrow [p]$, so that composition corresponds to composition of 
functions. A \emph{base morphism} of a theory $T$ is a morphism that arises
as the image of a morphism in the initial theory with respect to the unique theory
morphism $\Theta \rightarrow T$. For example, the base morphisms $n \rightarrow p$ 
in a theory $\Words_A$
are the morphisms of the form $(x_{1\rho},\ldots,x_{n \rho})$, 
where $\rho$ is a function $[n] \rightarrow [p]$. In any theory, we may represent 
base morphisms $n \rightarrow p$ as functions $[n]\rightarrow [p]$.

\section{Statement of the main result}

By taking sets of morphisms of a theory $T$, we may sometimes define a theory $P(T)$. 
(For a more general construction, the reader is referred to \cite{BEextension}.)

The morphisms $1 \rightarrow p$ in $P(T)$ are all sets $L \subseteq T(1,p)$. A morphism $n \rightarrow p$ is an $n$-tuple $(L_1,\ldots,L_n)$ 
of morphisms $1 \rightarrow p$, including the tuple $0_{n,p} = 
(\emptyset,\ldots,\emptyset)$. To define composition, suppose that 
$L: 1 \rightarrow p$ and $K = (K_1,\ldots,K_p): p \rightarrow q$. 
Then we define $L\cdot K : 1 \rightarrow q$ to be the set of all morphisms 
$1 \rightarrow q$ in $T$ of the form 
$$f\cdot \langle g_1,\ldots,g_m\rangle$$
such that $f: 1 \rightarrow m$ and there is a base morphism $\rho: m \rightarrow p$
with $f \cdot \rho \in L$ and $g_i \in K_{i \rho}$ for all $i \in [m]$.
When $L = (L_1,\ldots,L_n): n \rightarrow p$, we define $L \cdot K$
as the morphism $(L_1 \cdot K,\ldots,L_n \cdot K): n \rightarrow q$.  
For each $n\geq 0$, the identity morphism $\1{n}$ 
is the morphism $(\{1_n\},\ldots,\{n_n\}): n \rightarrow n$,
and the $i$th distinguished morphism $1 \to n$ is $\{i_n\}$. 

The above construction does not necessarily give a theory. When it 
does, we call $P(T)$ a \emph{power-set theory}. 

Suppose that $P(T)$ is a power-set theory.
We may equip $P(T)$ with a sum operation, denoted $+$ and defined by component wise set union. We define 
$$L + L' =  
(L_1\cup L_1',\ldots,L_n \cup L_n'): n \rightarrow p,$$
for all $L = (L_1,\ldots,L_n) : n \rightarrow p$ and $L' = (L_1',\ldots,L_n'): n \rightarrow p$ 
in $P(T)$. 
It is clear that, equipped with the operation $+$ and the constant $0_{n,p}$,
each hom-set $P(T)(n,p)$ is a commutative, idempotent monoid.
Moreover, 
\begin{eqnarray} 
\label{eq-grove1}
  i_n \cdot (L + L') &=& i_n \cdot L+ i_n \cdot L'\\
  \label{eq-grove2}
  i_n \cdot 0_{n,p} &=& 0_{1,p}\\
  \label{eq-grove3}
  (L + L') \cdot K  &=& L\cdot K + L'\cdot K\\
  \label{eq-grove4}
  0_{n,p}\cdot K &=& 0_{n,q},
\end{eqnarray} 
for all $L,L': n \rightarrow p$ and $K: p \rightarrow q$. 
(Here, we adapt the convention that composition has higher precedence than sum.)
Thus, $P(T)$ is an \emph{idempotent grove theory}
\cite{BEbook}. 

The above power-set construction is applicable to the theories $\Words_A$ and $\Tree_\Sigma$,
yielding the idempotent grove theories $\Lang_A$ of languages over $A$ and $\TreeLang_\Sigma$ 
of tree languages over $\Sigma$. 

Any power-set theory $P(T)$ is naturally equipped with a partial order $\subseteq$
defined by component-wise set inclusion. 
It is 
clear that each hom-set $P(T)(n,p)$ is a complete lattice with least element 
$0_{n,p}$, moreover, the theory operations are monotone, in fact continuous. 
(Composition preserves all suprema in its first argument, and  
tupling preserves all suprema in each of its arguments.) Thus,
we can define a \emph{dagger operation} $^\dagger: T(P)(n ,n + p) \rightarrow T(P)(n,p)$, $L \mapsto L^\dagger$,
by taking the least solution of the fixed point equation $X = L \cdot \langle X, \1{p}\rangle$. 
In particular, the theories $\Lang_A$ and $\TreeLang_\Sigma$ are also equipped with a 
dagger operation. The least subtheory of $\Lang_A$ containing the finite languages 
which is closed under dagger 
is the theory $\CFL_A$ of context-free languages, and the least subtheory of $\TreeLang_A$ 
containing the finite tree languages which is closed under dagger 
is the theory $\Reg_\Sigma$ of regular tree languages \cite{Esregtree1,Esregtree2,GecsegSteinby}.
Both $\CFL_A$ and $\Reg_\Sigma$ are idempotent grove theories.

We define yet another class of theories equipped with both an additive structure and a dagger 
operation, the theories of synchronization trees.
 A \emph{hyper tree} consists of a countable set $V$ of vertices and 
a countable set $E$ of edges, each edge $e$ having a source 
$v$ in $V$ and an ordered  sequence of target vertices 
$(v_1,\ldots,v_n) \in V^n$, for some $n \geq 0$. There is a 
distinguished vertex, the \emph{root} $v_0$, such that each vertex $v$ 
is the target vertex of a unique path from $v_0$ to $v$. 
An isomorphism between hyper trees is determined by a bijection 
between the vertices and a bijection between the edges that jointly  
preserve the root and the source and target of the edges.   

Synchronization trees over a set $A$ of \emph{action symbols} were defined in 
\cite{Winskel}.
A (slight) generalization of synchronization trees for ranked sets is given
in \cite{Essup}. Suppose that  $\Sigma$ is a ranked set.
A \emph{synchronization tree} $t = (V_t,E_t,\lambda_t): 1 \rightarrow p$ 
over $\Sigma$ is a hyper tree 
with vertex set $V_t$, hyper edges $E_t$, equipped with a labeling function 
$\lambda_t: E_t \rightarrow \Sigma \cup \{\ex_1,\ldots,\ex_p\}$,   
where the $\ex_i$ are referred to as the \emph{exit symbols}.  
Each hyper edge $e: v \rightarrow (v_1,\ldots,v_n)$ with source $v$ 
and target $(v_1,\ldots,v_n)$ is labeled in $\Sigma_n$,
or by an exit symbol when $n = 0$. When $t$ is a synchronization tree 
and $v$ is a vertex of $t$, then the vertices `accessible' from 
$v$ (including $v$) span the \emph{subtree} $t|_v$. The edges of $t|_v$ 
are those edges of $t$ having a source accessible from $v$.
Similarly, if $e$ is an edge $v \rightarrow (v_1,\ldots,v_n)$,
then $v$ and the vertices accessible from the $v_i$ determine 
a synchronization tree denoted $t|_e$ and called a \emph{summand} of $t|_v$. 
An isomorphism between synchronization trees is an isomorphism 
of the underlying hyper trees which preserves the labeling.
We usually identify isomorphic synchronization trees.  
A synchronization tree $n \rightarrow p$ over $\Sigma$ is an $n$-tuple 
$(t_1,\ldots,t_n)$ of synchronization trees $1 \rightarrow p$ over $\Sigma$. 
A synchronization tree $t: 1 \rightarrow p$ is \emph{finite} if its set of edges 
is finite (and thus its vertex set is also finite), 
\emph{finitely branching} if each vertex is the source of a 
finite number of edges, and \emph{regular}, 
if it has a finite number of subtrees (up to isomorphism)
and only a finite number of letters from $\Sigma$ appear as 
edge labels. When $t: 1 \to p$ is regular, the number of trees of the 
form $t|_e$ where $e$ is an edge is also finite. 
A synchronization tree $t: n \rightarrow p$ is finite (finitely branching, regular, resp.) 
if its components are all finite (finitely branching, regular, resp.).

We may identify each letter $\sigma \in \Sigma_n$ with the finite synchronization tree 
$1 \to n$ having and edge $r \to (v_1,\ldots,v_n)$ labeled $\sigma$,
where $r$ is the root,  and an edge originating in $v_i$ labeled $\ex_i$ 
for each $i \in [n]$. In the same way, we may view each exit symbol
$\ex_i$ as a tree $1 \to n$ for each $i \in [n]$, $n \geq 0$.

Synchronization trees over $\Sigma$ form a theory $\ST_\Sigma$. When $t: 1 \rightarrow p$
and $t' = (t_1',\ldots,t_p') : p \rightarrow q$, then $t\cdot t': 1 \rightarrow q$
is constructed from $t$ by replacing each edge of $t$ labeled $\ex_i$ 
for some $i\in [p]$ by a copy of $t_i'$. When $t= (t_1,\ldots,t_n): n \rightarrow p$,
then $t \cdot t' = (t_1\cdot t',\ldots,t_n \cdot t'): n \rightarrow q$. 
For each $i \in [n]$, the distinguished  morphism $i_n$ is the tree having a single edge labeled $\ex_i$.
For synchronization trees $t,t': 1 \rightarrow p$, we also define $t + t': 1 \rightarrow p$ 
as the tree obtained from (disjoint copies of) $t$ and $t'$ by merging the roots. When $t = (t_1,\ldots,t_n): n \to p$ 
and $t' = (t_1',\ldots,t_n'): n \rightarrow p$, then $t + t' = (t_1 + t_1', \ldots,
t_n + t_n'): n \to p$. We also define $0_{1,p}$ as the tree $1 \rightarrow p$ having no edge,
and $0_{n,p} = (0_{1,p},\ldots,0_{1,p}): n \rightarrow p$, for all $n,p \geq 0$.
 Clearly, each hom-set of $\ST_\Sigma$ is 
a commutative monoid and (\ref{eq-grove1})--(\ref{eq-grove4}) hold, so that 
$\ST_\Sigma$ is a \emph{grove theory} \cite{BEbook}.
We also define the grove theories $\FST_\Sigma$ of finite and $\RST_\Sigma$ 
of regular synchronization trees over $\Sigma$.

Suppose that $t$ and $t'$ are synchronization trees $1 \rightarrow p$ over $\Sigma$.
A \emph{simulation} \cite{Milner,Park} $t \rightarrow t'$ is a relation 
$R \subseteq V_{t} \times V_{t'}$, relating the roots 
such that whenever $e: v \rightarrow (v_1,\ldots,v_n)$ is an edge of $t$
and $v R v'$, then there is an equally labeled edge $v'\rightarrow (v_1',\ldots,v_n')$ of $t'$ 
such that $v_i R v_i'$ for all $i$. 
Note that the domain of a simulation $R: t \rightarrow t'$ is $V_t$. When $R$ 
is a simulation $t \rightarrow t'$ and $h$ is a function contained in $R$,
then $h$ is also a simulation, called a \emph{functional} simulation.
It is well-known 
that simulations compose, so that if $t,t',t'' : 1 \rightarrow p$ 
and $R$ is a simulation $t \rightarrow t'$ and $R'$ is a simulation $t' \rightarrow t''$,
then the relational composition of $R$ and $R'$ is a simulation $t\rightarrow t''$.
When $t=(t_1,\ldots,t_n)$ 
and $t' = (t_1',\ldots,t_n')$ are synchronization trees $n \rightarrow p$,
a simulation $t \rightarrow t'$ is an $n$-tuple $(R_1,\ldots,R_n)$, where 
each $R_i$ is a simulation $t_i \rightarrow t_i'$. 
We say that $t$ and $t'$ are \emph{simulation equivalent}, 
denoted $t \equiv_s t'$, if there are simulations $t \rightarrow t'$ and 
$t' \rightarrow t$. 
The relation $\equiv_s$ is a grove theory congruence of $\ST_\Sigma$, i.e., 
a theory congruence which preserves the sum operation, giving rise to the 
grove theory  $\SST_\Sigma = \ST_\Sigma / \equiv_s$. 
We will denote the simulation equivalence class of a tree $t$ by $[t]_s$,
or sometimes just $[t]$. Moreover, when $t = (t_1,\ldots,t_n): n \to p$, we 
identify $[t]_s$ with $([t_1]_s,\ldots,[t_n]_s)$. 

We define the relation $t \sqsubseteq_s t'$ for synchronization trees $t,t' : n \rightarrow p$ 
iff there is a simulation $t \rightarrow t'$. Also, we define $[t]_s \sqsubseteq_s [t']_s$ 
iff $t \sqsubseteq_s t'$, since the definition is independent of the choice 
of the representatives of the equivalence classes. Since simulations compose,
the relation $\sqsubseteq_s$ is a pre-order on synchronization trees 
and a partial order on simulation equivalence classes. Each 
hom-set of $\SST_\Sigma$ has all countable suprema. Indeed, when $t_i,\ i \in I$,  
is a countable family of trees $n \rightarrow p$, then $\sup_{i\in I} [t_i] = [t]$ 
for the tree $t = \sum_{i \in I}t_i : n \rightarrow p$ obtained by taking 
the disjoint union of the $t_i$ and identifying the roots. 
When $I$ is empty, the sum is the tree $0_{n,p}$. The theory operations 
are $\omega$-continuous, so that we can define a dagger operation. For each 
$f =[t]_s: n \rightarrow n + p$ in $\SST_\Sigma$, $f^\dagger: n \rightarrow p$ is 
 the least solution of the fixed-point equation $x = f \cdot \langle x,\1{p}\rangle$. 
The least subtheory of $\SST_\Sigma$ containing the finite synchronization trees 
which is closed under dagger is the theory $\SRST_\Sigma$ of simulation equivalence 
classes containing at least one regular tree. Further, we denote by $\SFST_\Sigma$ 
the subtheory determined by those simulation equivalence classes containing at 
least one finite synchronization tree. Both $\SRST_\Sigma$ and $\SFST_\Sigma$ 
are closed under the sum operation and both of them are grove theories. Note that 
we may identify $\SRST_\Sigma$ with $\RST_\Sigma/\equiv_s$ 
and $\SFST_\Sigma$ with $\FST_\Sigma/ \equiv_s$.

A \emph{term} is well-formed expression composed of morphism variables 
and constants for the distinguished morphisms using the theory operations,
sum, and dagger. Each term has a source $n$ and a target $p$,
for some nonnegative integers $n,p$.  

We are now ready to state our main result. We may view each set $A$ 
as a ranked set where each letter has rank $1$.

\begin{thm}
\label{thm-main}
The following conditions are equivalent for sorted terms $t,t': n \rightarrow p$ 
constructed from sorted morphism variables by the theory operations, 
sum, and dagger. 
\begin{itemize}
\item The identity $t = t'$ holds in all power-set theories.
\item The identity $t = t'$ holds in all theories $\Lang_A$ (or $\CFL_A$), where $A$ is a set.
\item The identity $t = t'$ holds in all theories $\TreeLang_\Sigma$ (or $\Reg_\Sigma$), where $\Sigma$ is a ranked set.
\item The identity $t = t'$ holds in all theories $\SST_\Sigma$ (or $\SRST_\Sigma$), where $\Sigma$ is a ranked set.
\item The identity $t = t'$ holds in all theories $\SST_A$ (or $\SRST_A$), where $A$ is a set.
\end{itemize}
\end{thm}

(In the second and fifth items, we could as well require that $A$ is a two-element set.) 
The proof of Theorem~\ref{thm-main} will be completed in Section~\ref{sec-main-proof}.

Since simulation equivalence is known to be decidable (in polynomial time 
for finite process graphs, cf. \cite{BGS92,JS01}), it follows that it is decidable for terms 
$t,t':n \rightarrow p$ whether $t = t'$ holds in 
in all theories $\CFL_A$. This fact is in contrast with the well-known undecidability
of the equivalence problem for context-free grammars. Intuitively, our positive result 
is due to the fact that we are interested in the equivalence of terms under \emph{all} 
possible interpretations of the morphism variables as context-free languages, and we do not have 
a constant for the language $\{x_1x_2\}:1 \rightarrow 2$. By adding this constant, we 
would run into 
undecidability.

\begin{remark}
Languages and tree languages satisfy 
\begin{eqnarray*}
L \cdot \langle L_1 + L_1',\ldots,L_n + L_n'\rangle &=& \sum_{K_i \in \{L_i,L_i'\}}
L \cdot \langle K_1,\ldots,K_n\rangle\\
L\cdot \langle L_1,\ldots,0_{1,p},\ldots,L_n\rangle &=& 0_{1,p}
\end{eqnarray*}
for all $L: 1 \rightarrow n$, and $L_i,L_i': 1 \rightarrow p$ whenever each of the variables
$x_1,\ldots,x_n$ occurs \emph{exactly once} in each word/tree of $L$. However,
these equations do not hold universally. 
\end{remark}

\section{Free $\omega$-continuous idempotent grove theories}

Recall that a grove theory is theory $T$ with a commutative 
additive monoid structure $(T(n,p),+,0_{n,p})$ on each hom-set
such that (\ref{eq-grove1})--(\ref{eq-grove4}) hold. 
An idempotent grove theory is a grove theory with an idempotent
sum operation. A morphism of (idempotent) grove theories is a theory 
morphism preserving $+$ and the constants $0_{n,p}$. When $T$ is an idempotent 
grove theory, we may define a partial order $\leq$ on each hom-set 
$T(n,p)$ by $f \leq g$ iff $f + g = g$. It is clear that 
$0_{n,p}$ is the least element of $T(n,p)$ with respect to this partial order,
and the tupling and sum operations preserve the order.
Composition necessarily preserves the order 
in the first argument, but not necessarily in the second.
When it does, we call $T$ an \emph{ordered idempotent grove theory}.
Moreover, when $f,g: n \rightarrow p$, then $f \leq g$ iff $i_n \cdot f \leq i_n \cdot g$ 
for all $i \in [n]$. Thus, the partial order on morphisms $n \rightarrow p$
is determined by the order on the morphisms $1 \rightarrow p$.
Morphisms of idempotent grove theories necessarily preserve the order.

We say that an idempotent grove theory is \emph{$\omega$-continuous} if the supremum 
$\sup_k f_k$ of each $\omega$-chain $(f_k: n \rightarrow p)_k$ exists and composition 
preserves the supremum of $\omega$-chains in both arguments.
It follows that every $\omega$-continuous idempotent 
grove theory is ordered, and the supremum of 
every countable family of morphisms $f_i: n \rightarrow p$, $i \in I$ exists.
Moreover, composition preserves the supremum of all countable families in 
its first argument.  A morphism of $\omega$-continuous idempotent grove theories 
preserves the supremum of $\omega$-chains.

Examples of $\omega$-continuous idempotent grove theories include the 
theories $\Lang_A$, $\TreeLang_\Sigma$ and $\SST_\Sigma$ defined above. 
In $\Lang_A$ and $\TreeLang_A$, the relation $\leq$ is the component-wise  
set inclusion relation $\subseteq$, whereas it is the relation $\sqsubseteq_s$ in $\SST_A$.
Each of these theories is equipped with a dagger operation. More generally, 
we may define a dagger operation in any $\omega$-continuous idempotent grove theory:
 for a morphism $f:n \rightarrow n + p$,
$f^\dagger: n\rightarrow p$ is the least solution of the equation $x = f \cdot \langle x,\1{p}\rangle$. We have $f^\dagger = \sup_k f^{(k)}$, where $f^{(0)} = 0_{n,p}$ and 
$f^{(k+1)} = f \cdot \langle f^{(k)},\1{p}\rangle$, for all $k \geq 0$. 
It is clear that every morphism of $\omega$-continuous idempotent 
grove theories preserves dagger.

In this section, our aim is to give a concrete description of the free $\omega$-continuous idempotent grove theories and relate them to the 
theories $\SST_\Sigma$. 
To this end, we will need some more facts about synchronization trees. 
Suppose that $t,t': 1 \rightarrow p$ in $\ST_\Sigma$.
A \emph{morphism} $t \rightarrow t'$ is a function $\tau: V_t \rightarrow V_{t'}$ that preserves the root and such that if there is an edge $v \rightarrow (v_1,\ldots,v_n)$ in $t$,
then there is an equally labeled edge $v\varphi \rightarrow (v_1\varphi,\ldots,v_n\varphi)$ 
in $t'$. Thus, we may extend $\tau$ to a function $E_t \rightarrow E_{t'}$. 
Note that a morphism is just a functional simulation, so that 
$t\sqsubseteq_s t'$ for $t,t' : 1 \rightarrow p$ in $\ST_\Sigma$ iff
there is a morphism $t \rightarrow t'$.
 Morphisms between trees $1 \rightarrow p$ compose by function
composition; we denote the composite of $\rho: t \rightarrow t'$ and $\tau: t' \rightarrow t''$ 
by $\rho \star \tau$. For each $t: 1\rightarrow p$, there is an identity morphism $I(t): t \rightarrow t$.
When $t,t': n \rightarrow p$ with $t= (t_1,\ldots,t_n)$ and 
$t' = (t'_1,\ldots,t'_n)$, then a morphism $t \rightarrow t'$ is an $n$-tuple 
$\rho =(\rho_1,\ldots,\rho_n)$ of morphisms $\rho_i: t_i \rightarrow t_i'$, $i \in [n]$.
When $\tau = (\tau_1,\ldots,\tau_n): t' \rightarrow t''$ is another morphism, 
where $t'' = (t_1'',\ldots,t_n''): n \rightarrow p$, $\rho\star \tau$ is 
$(\rho_1\star \tau_1,\ldots,\rho_n\star \tau_n)$. Composition is associative, 
and the morphisms $I(t) = (I(t_1),\ldots,I(t_n))$ act as identities. 
Thus, each hom-set $T(n,p)$ is a category, called a \emph{vertical} category.
It is shown in \cite{BEbook} that each vertical category $\ST_\Sigma(n,p)$ 
is countably cocomplete.

We define another, `horizontal' composition of synchronization tree 
morphisms. Suppose that $t,t': 1 \rightarrow p$ and $s,s': p \rightarrow q$ with components 
$s_i,s'_i: 1 \rightarrow q$, $i \in [p]$. Let $\rho: t \rightarrow t'$ and 
$\tau = (\tau_1,\ldots,\tau_p): s \rightarrow s'$. Recall that $t \cdot s$ is the tree 
obtained by replacing each exit edge of $t$ labeled $\ex_i$ for some $i\in [p]$
by a copy of $s_i$, and similarly for $t'\cdot s'$. Now we define 
$\rho \cdot \tau$ to be the function that acts as $\rho$ on the vertices 
of $t$ and as $\tau_i$ on those vertices of copies of each $s_i$ that replace
an exit edge 
of $t$ labeled $\ex_i$. When $t,t': n \rightarrow p$ and $\rho: t \rightarrow t'$, we define 
$\rho \cdot \tau: t\cdot s \rightarrow t'\cdot s'$ component wise. It is clear 
that the `interchange law' holds:
\begin{eqnarray*}
(\rho\star \rho')\cdot (\tau \star \tau') &=& (\rho \star \tau) \cdot (\rho' \star \tau'),
\end{eqnarray*}
for all appropriate morphisms $\rho,\rho',\tau,\tau'$. 
Also, $I(t) \cdot I(t') = I(t \cdot t')$ 
for all $t: n \to p$ and $t': p \to q$.
It is shown in \cite{BEbook}
that horizontal composition preserves 
colimits of $\omega$-diagrams. 
We can also define the sum $\rho + \tau: t + s \to t' + s'$ of synchronization 
tree morphisms $\rho: t \to t'$ and $\tau: s \to s'$, $t,t',s,s': 1 \to p$ so 
that it acts as $\tau$ on the vertices of $t$ and as $\rho$ on the vertices of $s$.
When $\rho = (\rho_1,\ldots,\rho_n)$ and $\tau = (\tau_1,\ldots,\tau_n)$ are appropriate
synchronization tree morphisms between synchronization trees $n \to p$, 
we define $\rho + \tau = (\rho_1+ \tau_1,\ldots,\rho_n + \tau_n)$.

It is known that each fixed point equation $x = f \cdot \langle x,\1{p}\rangle $ 
for $f: n \to n + p$ in $\ST_\Sigma$ 
has an `\emph{initial}' solution obtained as the colimit of the
$\omega$-diagram $(\tau_k: f^{(k)} \to f^{(k+1)})_k$. Here, we define $f^{(0)} = 
0_{n,p}$ and $f^{(k+1)} = f \cdot \langle f^{(k)},\1{p}\rangle$ for all $k \geq 0$.
Further, $\tau_0$ is the unique morphism $f^{(0)} \to f^{(1)}$, and 
$\tau_{k+1} = f \cdot \langle \tau_k,\1{p}\rangle$ for all $k \geq 0$.
Denoting this initial solution by $f^\dagger$, we obtain a dagger operation 
$^\dagger: \ST_\Sigma(n,n+p) \to \ST_\Sigma(n,p)$, $n,p \geq 0$. Since
the morphisms $\tau_k$ are injective, we may suppose that each $\tau_k$ 
is in fact the inclusion of $f^{(k)}$ into $f^{(k+1)}$. The colimit 
$f^\dagger$ then becomes the union of the $f^{(k)}$. 
It is known that when $f = (f_1,\ldots,f_n) : n \to n + p$ 
is `guarded' in the sense that  
none of the $f_i$ has an exit edge labeled $\ex_{j}$ with $j \in [n]$ 
originating in the root, then $f^\dagger$ is the unique solution (up to isomorphism) 
of the fixed point equation for $f$. Moreover, the least subtheory of 
$\ST_\Sigma$ with respect to set inclusion which contains the finite trees 
and is closed under dagger is $\RST_\Sigma$, the theory of regular
synchronization trees.

An \emph{ideal} in $\FST_\Sigma(n,p)$ is a nonempty set $Q \subseteq \FST_\Sigma$
which is downward closed with respect to the relation $\sqsubseteq_s$. An 
\emph{$\omega$-ideal} is an ideal $Q$ which is generated by some $\omega$-chain 
$(t_k)_k$ of trees $t_k: n \rightarrow p$ in $\FST_\Sigma$
with $t_k \sqsubseteq_s t_{k+1}$ for all $k \geq 0$. Note that we may 
identify any ($\omega$)-ideal $Q \subseteq \FST_\Sigma(n,p)$  with an $n$-tuple 
of ($\omega$)-ideals $(Q_1,\ldots,Q_n)$, where $Q_i \subseteq \FST_\Sigma(1,p)$ 
is the set of all $i$th components of the members of $Q$, for each $i \in [n]$. 
We may recover $Q$ from $(Q_1,\ldots,Q_n)$ as the set 
$\{t : n \rightarrow p: i_n \cdot t \in Q_i\ {\rm for}\ {\rm all}\ i \in [n]\}$.

We may turn $\omega$-ideals into an idempotent 
grove theory $\omegaSFST_\Sigma$. The set of morphisms $n \to p$
in $\omegaSFST_\Sigma$ is the collection of all $\omega$-ideals 
$Q \subseteq \FST_\Sigma(n,p)$.
When $Q:n \rightarrow p$ and $Q' : p\rightarrow q$, then we define 
$Q\cdot Q' : n \rightarrow q$ to be the ideal generated 
by the set of all trees $f\cdot g$ with $f: n \rightarrow p$ in $Q$ 
and $g: p \rightarrow q$ in $Q'$. When $Q$ and $Q'$ are generated by the 
$\omega$-chains $(f_k)_k$ and $(g_k)_k$, then $Q\cdot Q'$ 
is the $\omega$-ideal generated  by $(f_k \cdot g_k)_k$.
For each $i \in [n]$, $n \geq 0$, the distinguished morphism 
$1 \rightarrow n$ is the ideal generated by the tree $\ex_i$.
 The sum $Q +Q': n \to p$ of $Q: n \rightarrow p$ and $Q': n\rightarrow p$ 
is defined as the ideal generated by $\{f + g : f \in Q,\ g\in Q'\}$.
It is easy to see that this is again an $\omega$-ideal.
The morphism $0_{n,p}: n \rightarrow p$ in $\omegaSFST_\Sigma$ 
is the ideal containing only the tree $0_{n,p}$.

There is a canonical embedding of $\SFST_\Sigma$ into $\omegaSFST_\Sigma$
which maps the simulation equivalence class of a finite tree $t: n \to p$ to the principal $\omega$-ideal 
$(t] = \{t': n \to p : t' \sqsubseteq_s t\}$. It is easy to see that 
this defines an ordered idempotent grove theory morphism 
$\SFST_\Sigma \to \omegaSFST_\Sigma$.

An $\omega$-ideal in $\SFST_\Sigma(n,p)$ is defined in the same 
way as in $\FST_\Sigma(n,p)$ using the partial order $\sqsubseteq_s$. 
We may identify any $\omega$-ideal $Q \subseteq \SFST_\Sigma(n,p)$ 
with an $\omega$-ideal in $Q' \subseteq  \FST_\Sigma(n,p)$ 
which is the union  
of all simulation equivalence classes of the trees in $Q$.
Using this identification, $\omegaSFST_\Sigma$ is just the 
completion of $\SFST_\Sigma$ by $\omega$-ideals as defined 
in \cite{Bloomordered}. It follows from the main result
of \cite{Bloomordered} that $\omegaSFST_\Sigma$ is an 
$\omega$-continuous idempotent grove theory, and we that have:

\begin{prop}
\label{prop-bloom}
The theory $\omegaSFST_\Sigma$ is the free $\omega$-continuous idempotent grove theory, 
freely generated by $\SFST_\Sigma$. Given any $\omega$-continuous idempotent grove 
theory $T$ and an ordered idempotent grove theory morphism $\varphi: \SFST_\Sigma \to T$,
there is a unique $\omega$-idempotent grove theory morphism 
$\varphi^\sharp : \omegaSFST_\Sigma \to T$ extending $\varphi$. 
\end{prop}

\begin{prop}
\label{prop-free-finite}
The theory $\SFST_\Sigma$ is the free ordered idempotent grove theory, 
freely generated by $\Sigma$.
\end{prop}

{\sl Proof.} It is known that $\FST_\Sigma$ is the free grove theory, 
freely generated
by $\Sigma$, cf. \cite{BEbook}. Let $\approx$ denote the least grove theory 
congruence such that $\FST_\Sigma/\approx$ is an ordered idempotent grove theory,
and define $f \preceq g$ iff $f + g \approx g$, for all $f,g: n \to p$. 
Thus, $f \approx g$ iff both $f \preceq g$ and $g\preceq f$ hold. 
We show that the relations $\equiv_s$ and $\approx$ are equal. The inclusion of 
$\approx$ is $\equiv_s$ is clear, since $\SFST_\Sigma$ is an ordered 
idempotent grove theory. To complete the proof, we show that 
for all $f,g: 1 \to p$ in $\FST_\Sigma$, if $f \sqsubseteq g$,
then $f \preceq g$. We argue by induction on the height\footnote{The height is 
the length of the longest path.}
of $f$.
When the height of $f$ is $0$, $f = 0_{1,p}$ and our claim is 
clear. Suppose now that the height of $f$ is positive.
If the root of $f$ is the source of a single edge, then 
$f = \sigma \cdot \langle f'_1,\ldots,f'_k \rangle$ or $f = j_p$ 
for some $\sigma \in \Sigma_k$, $f'_1,\ldots,f'_k: 1 \rightarrow p$ 
and $j \in [p]$. In the first case, since $f \sqsubseteq g$,
we may write $g$ as $g_0 + \sigma \cdot \langle g'_1,\ldots,g'_k\rangle$
for some $g_0,g'_1,\ldots,g'_k : 1 \rightarrow p$ with $f'_i \sqsubseteq_s g'_i$
for all $i \in [k]$. By the induction hypothesis, we have 
$f'_i \preceq g'_i$ for all $i \in [k]$ and thus $f\preceq g$.
In the second case, when $f = j_p$, $g$ can be written as $g_0 + j_p$,
for some $g_0: 1 \rightarrow p$. Thus, $f \preceq g$ again. 
Suppose finally that the root of $f$ has 2 or more outgoing edges $e$.
Then we van write $f$ as a finite sum of summands $f|_e$. By the previous case 
and the induction hypothesis we have $f_e \preceq g$ for each $e$. 
Since sum is idempotent, we conclude that $f \preceq g$. \eop

By Proposition~\ref{prop-free-finite} and Proposition~\ref{prop-bloom}, we immediately have:

\begin{thm}
\label{thm-free-omegaideal}
For each ranked alphabet $\Sigma$,
the theory $\omegaSFST_\Sigma$ is the 
free $\omega$-continuous idempotent grove 
theory, freely generated by $\Sigma$.
\end{thm}

Our next task is to relate $\omega$-ideals of finite synchronization trees to 
possibly infinite synchronization trees. By the \emph{prefix} of a tree $t: 1 \rightarrow p$ 
of height at most $n \geq 0$ we shall mean the tree $t_n: 1 \rightarrow p$ 
obtained from $t$ by removing all vertices at distance greater than $n$ 
from the root together with all incident edges.

We call a synchronization tree $t:1 \rightarrow p$ \emph{(simulation) reduced} 
if whenever $v$ is vertex of $t$ and $e$ and $e'$ are different
edges originating in $v$, then the trees $t|_e$ and $t|_{e'}$ are \emph{not}
comparable by the relation $\sqsubseteq_s$. Not all simulation equivalence classes 
of synchronization trees contain a reduced tree. For example, 
suppose that $\sigma\in \Sigma_1$, and let 
$t= \sum_{n \geq 0}\sigma^n \cdot 0: 1 \rightarrow 0$. Then the 
simulation equivalence class of $t$ does not contain a reduced tree.
However, each finitely branching tree $t: 1 \rightarrow p$ 
is simulation equivalent to a (finitely branching) reduced tree $t': 1 \rightarrow p$.
The tree $t'$ can be constructed from $t$ by starting at the root
and at each vertex $v$, keeping only one summand from each maximal
simulation equivalence class of the trees $t|_{e}$, where $e$ ranges over the
set of edges originating in $v$. Note that when $t$ 
is reduced, then so is any subtree $t|_v$, where $v$ is a vertex of $t$.
It follows that if the simulation
equivalence class of a tree contains a finitely branching tree, 
then it also contains a reduced tree. In particular, each regular tree 
is simulation equivalent to a reduced (regular) tree. When $t$ is reduced, 
then $t$ is the unique reduced tree in its $\equiv_s$-equivalence class $[t]_s$.

\begin{lem}
A synchronization tree $t: 1 \rightarrow p$ in $\ST_\Sigma$ is reduced 
iff the single morphism $t \rightarrow t$ is the identity morphism.
\end{lem}

{\sl Proof.} Suppose first that $t$ is reduced, and let $\tau$ be a
a morphism $t \rightarrow t$. Let $v$ be a vertex of $t$ of distance $n$ 
from the root. We show by induction on $n$ that $v\tau = v$.
When $n = 0$, this is clear, since both $v$ and $v\tau$ are 
equal to the root of $v$. Suppose now that $n > 0$ and that 
we have proved our claim for vertices of distance less than $n$ 
from the root. Consider the edge $e: u \rightarrow (v_1,\ldots,v_k)$ of $f$ 
such that $v$ is one of the $v_i$ for some $i \in [k]$. Since $u\tau = u$ and since 
$t|_e$ is not comparable with respect to $\sqsubseteq_s$ 
to any other summand $t|_e$ of $t|_u$, 
we must have $v_i\tau = v_i$ for all $i$ 
and thus $v\tau = v$.

Suppose now that $t$ is not reduced. Then there is some vertex 
$v$ with different outgoing edges $e,e'$ such that $t|_e 
\sqsubseteq_s t|_{e'}$. Let $\rho$ denote a morphism $t|_e \rightarrow t_|{e'}$.
Using $\rho$, we may construct a nontrivial morphism $\tau: t \rightarrow t$.
We define $u\tau = u$ if $u$ is not a vertex of $t|_e$,
and $u\tau =u\rho$ otherwise. \eop

\begin{cor}
Suppose that $t,t': 1 \rightarrow p$ are simulation equivalent 
reduced trees in $\ST_\Sigma$. Then $t$ is isomorphic 
to $t'$.
\end{cor}

We will call a tree $t = (t_1,\ldots,t_n): n \rightarrow p$ (simulation) reduced if its 
components $t_i$ are. The above facts also apply for 
reduced trees $t: n \rightarrow p$.


For each tree $t : n \rightarrow p$ in $\ST_\Sigma$, 
let $K(t)$ denote the set of all \emph{finite} trees 
$t': n \rightarrow p$ with $t' \sqsubseteq_s t$. 

\begin{prop}
A set of finite trees $Q \subseteq \FST_\Sigma(n,p)$ is an 
$\omega$-ideal iff $Q = K(t)$ for some (possibly infinite) tree 
$t: n \rightarrow p$ in $\ST_\Sigma$.
\end{prop}

{\sl Proof.} 
It suffices to prove the claim for $n = 1$.
Suppose first that $Q = K(t)$ for some $t:1 \rightarrow p$ in $\ST_\Sigma$.
Then $Q$ is the $\omega$-ideal generated by the $\omega$-chain 
$(t_k)_k$, where $t_k$ is the prefix of $t$ of height at most $k$.

Suppose now that $Q$ is the $\omega$-ideal generated by the 
$\omega$-chain $(t_k)_k$ of finite trees $1 \rightarrow p$. 
Let $t = \sum_{k \geq 0}t_k$. Then for any finite tree $s: 1 \rightarrow p$,
$s \sqsubseteq_s t$ iff $s \sqsubseteq_s \sum_{k = 0}^n t_k$ 
iff $s \sqsubseteq_s t_k$ for some $k \geq 0$. \eop

\begin{prop}
Suppose that $t,t': n \rightarrow p$ in $\ST_\Sigma$. If $t \sqsubseteq_s t'$ 
then $K(t) \subseteq K(t')$. Moreover, if $t$ and $t'$ are finitely 
branching, or simulation equivalent 
to some finitely branching trees,
 and if $K(t) \subseteq K(t')$, then $t \sqsubseteq_s t$.
\end{prop}

{\sl Proof.} The first statement is obvious.
In order to prove the second, we may restrict ourselves 
to finitely branching trees $1 \rightarrow p$. So suppose that $t,t': 1 \rightarrow p$ are finitely 
branching with $K(t) \subseteq K(t')$. 
For each $k \geq 0$, let $t_k$ and $t'_k$ denote the (finite) 
prefixes of $t$ and $t'$ of height at most $k$, so that $t$ 
is the union of the $t_k$ and, similarly, $t'$ is the union of the $t'_k$,
for $k \geq 0$. Since $K(t) \subseteq K(t')$, we have $t_k \sqsubseteq_s t'_k$ 
for each $k \geq 0$, so that there is a finite nonempty set of morphisms 
$t_k \rightarrow t'_k$ for each $k\geq 0$. Also, the restriction of a morphism $t_{k+1}\rightarrow 
t'_{k+1}$ onto $t+k$ is a morphism $t_k \rightarrow t'_k$. By K\"onig's lemma, we may 
select a sequence of morphisms  $(\tau_k: t_k \rightarrow t_{k+1})_k$ 
such that $\tau_k$ is the restriction of $\tau_{k+1}$ for each $k \geq 0$.
Since $t$ is the union of the $t_k$ and $t'$ is the union of the $t'_k$,
the sequence $(\tau_k)_k$ determines a morphism $\tau: t \rightarrow t'$.
\eop

\begin{cor}
If $t,t': n \rightarrow p$ in $\ST_\Sigma$ are  
simulation equivalent to finitely branching trees, 
then $t \sqsubseteq_s t'$ iff $K(t) \subseteq K(t')$, 
 and $t \equiv_s t'$ iff $K(t) = K(t')$. 
\end{cor}

\begin{expl}
Let $t$ be the tree $t= \sum_{n \geq 0}\sigma^n \cdot 0_{1,0}: 1 \rightarrow 0$,
 and let $t' = \sigma^\omega: 1 \rightarrow 0$, a tree 
consisting of a single infinite branch with edges labeled $\sigma \in \Sigma_1$.
Then $K(t) = K(t')$ but $t \equiv_s t'$ does not hold. 
\end{expl}

Since every regular synchronization tree is simulation equivalent to 
a finitely branching regular tree, we have:

\begin{cor}
\label{cor-pr}
Suppose that $t,t':n \rightarrow p$ in $\RST_\Sigma$. 
Then $t \sqsubseteq_s t'$ iff $K(t) \subseteq K(t')$
 and $t \equiv_s t'$ iff $K(t) = K(t')$. 
\end{cor}

From Theorem~\ref{thm-free-omegaideal} and Corollary~\ref{cor-pr}, we obtain:

\begin{cor}
\label{cor-free}
Suppose that $\Sigma$ is a ranked set, $T$ is an $\omega$-continuous idempotent 
grove theory and $\varphi: \Sigma \rightarrow T$ is a rank preserving function. 
Then there is a unique way to extend $\varphi$ to an idempotent 
grove theory morphism $\varphi^\sharp : \SRST_\Sigma \rightarrow T$
preserving dagger.
\end{cor}

{\sl Proof.} 
Suppose that $T$ is an $\omega$-continuous idempotent grove theory and 
$\varphi$ is a rank preserving function $\Sigma \rightarrow T$. 
We may extend $\varphi$ to a morphism $\psi: \omegaSFST_\Sigma \rightarrow T$
of $\omega$-continuous idempotent grove theories. We know that 
$\SRST_\Sigma$ embeds in $\omegaSFST_\Sigma$ by the function which
maps a regular tree $t: n \rightarrow p$ to $K(t)$. It is a routine matter 
to verify that the embedding preserves the theory operations, the additive structure,
and dagger. Thus, we may identify $\SRST_\Sigma$ with a subtheory of 
$\omegaSFST_\Sigma$. The restriction of $\psi$ to $\SRST_\Sigma$ 
is the required extension $\varphi^\sharp: \SRST_\Sigma \rightarrow T$.
\eop

\section{Proof of the main result}
\label{sec-main-proof}

In this section our aim is to prove Theorem~\ref{thm-main}.

Recall that we may view each set $A$ as a ranked set of
letters of rank 1. We start by showing that for each 
ranked set $\Sigma$ there is some set $A$ such that 
$\SST_\Sigma$ embeds in $\SST_A$ and $\SRST_\Sigma$ embeds in $\SRST_A$.

\begin{prop}
\label{prop-emb0}
For every ranked set $\Sigma$ there exist a set $A$ 
and an injective (idempotent) grove theory 
morphism $\SRST_\Sigma \rightarrow \SRST_A$
preserving dagger.
\end{prop} 
 
{\sl Proof.} 
When $\Sigma$ is a ranked set, define 
$A = \ol{\Sigma} \cup \{\#\},$ where $\ol{\Sigma} = \{\ol{\sigma} :\sigma \in \Sigma\}$.
Our aim is to show that there is an injective dagger preserving 
grove theory morphism 
$\SRST_\Sigma \rightarrow \SRST_A$.

Consider the function $\varphi$ which maps the simulation equivalence 
class of the tree corresponding to a letter $\sigma \in \Sigma_n$, 
$n \geq 0$,  to the simulation equivalence class of the synchronization tree
\begin{eqnarray*}
s_\sigma = \ol{\sigma} \cdot (\# \cdot 1_n + \#^2 \cdot 2_n + \ldots + 
 \#^n \cdot n_n) : 1 \rightarrow n
 \end{eqnarray*}
  in $\ST_A$. (Recall that $\ol{\sigma}$ has rank 1. The tree $s_\sigma$ 
  has a single edge originating in the root, which is labeled 
  $\ol{\sigma}$.
  When $n = 0$, the simulation equivalence class of the tree $\sigma$ 
  is mapped to the equivalence class $[\ol{\sigma} \cdot 0_{1,0}]_s$.
By Corollary~\ref{cor-free}, 
this function can be extended in a unique way to an idempotent 
grove theory morphism $\varphi: \SRST_\Sigma \rightarrow \SRST_A$
preserving dagger.

It is not hard to see that $\varphi$ takes the following concrete form.
Suppose that $t: 1 \rightarrow p$ in $\RST_\Sigma$. Then $[t]_s\varphi$ 
is the equivalence class of the (regular) tree 
$t': 1 \rightarrow p$ in $\RST_A$ obtained from $t$ by replacing each edge labeled $\sigma\in \Sigma$ 
by a copy of the tree $s_\sigma$. Formally,
the set of vertices of $t'$  consists of the vertices of $t$ together 
with a vertex 
$[v,(v_1,v_2, \ldots,v_n)]$ and vertices $(v_i,j)$ 
with $1 < j < i \leq n$, for each hyper-edge $v \rightarrow (v_1,\ldots,v_n)$
of $t$.

The edges of $t'$ are the exit edges of $t$ labeled 
as in $t$ together with the following ones, where we suppose that 
$e: v\rightarrow (v_1,\ldots,v_n)$ is a hyper-edge of $t$ labeled $\sigma$.
\begin{enumerate}
\item An edge $v \rightarrow [v, (v_1,\ldots,v_n)]$ labeled 
      $\ol{\sigma}$.
\item An edge $[v,(v_1,\ldots,v_n)] \rightarrow (v_i,1)$
      for each $1 < i \leq n$ labeled $\#$.
\item An edge $(v_i,j) \rightarrow (v_i,j+1)$ and an edge 
      $(v_i,i-1) \rightarrow v_i$ labeled $\#$, for all $1< i \leq n$ 
      and $1 < j < i -1$.
\item An edge  $[v,(v_1,\ldots,v_n)] \rightarrow v_1$ labeled $\#$.
\end{enumerate}
When $t = (t_1,\ldots,t_n): n \rightarrow p$ and each $[t_i]_s$ 
is mapped to $[t'_i]_s$, then $[t]_s \varphi = ([t_1']_s,\ldots,[t_n']_s)$. 
Since each $t:1 \rightarrow p$ can be recovered from $t': 1\rightarrow p$, 
$[t]_s$ is uniquely determined by $[t']_s$, i.e., 
$\varphi$ is injective. 
 \eop 

\begin{remark}
The above proof can be extended to all synchronization trees to obtain an 
injective continuous idempotent grove theory morphism
$\SST_\Sigma \rightarrow \SST_A$.
\end{remark}

\begin{prop}
\label{prop-emb1}
For each set $A$ there exist a set $B$ and an injective 
dagger preserving (idempotent) grove theory morphism 
$\SRST_A \rightarrow \CFL_B$.
\end{prop}

{\sl Proof.}
Let $B = A \cup \{\#,\$\}$ and consider the dagger preserving 
grove theory morphism $\varphi: \SRST_A \rightarrow \CFL_B$
defined by the assignment 
$a \mapsto a(\#x_1\$)^* = \{a,a\#x_1\$,a(\#x_1\$)^2,\ldots\}
: 1 \rightarrow 1$, so that the image of each letter $a\in A$ is a regular language.
We claim that for any regular trees $t,s: 1 \rightarrow p$ in $\RST_A$, 
$$[t]\sqsubseteq_s [s] \Leftrightarrow  [t]\varphi \subseteq [s]\varphi.$$
The implication from left-to-right is immediate from Corollary~\ref{cor-free}.  

Suppose now that $t \not \sqsubseteq_s s$. We want to prove that 
$[t]\varphi \not\subseteq [s]\varphi$. We consider only the case $p = 0$
since the argument is similar for $p > 0$. 

The $n$-round \emph{simulation game} on the pair $(t,s)$ is played by two players, 
player I and II. In each round, player I selects an edge originating
in the vertex of $t$ entered in the previous round, 
or in the root in 
the first round, and player II must respond by selecting an equally 
labeled edge 
originating in the vertex of $s$ entered in the previous round, 
or in the root of $s$ in first round. Player I wins the play if player II
cannot respond. Otherwise player II wins. 
Since $t \not \sqsubseteq_s s$, there is some $n \geq 1$ such that player 
I wins the $n$-round simulation game on $(t,s)$. 
We show by induction on $n$ that $[t]\varphi \not \subseteq [s]\varphi$. 
When $n = 1$, player I can choose an edge originating in the root 
of $t$ whose label is not matched by the label of any edge 
originating in the root of $s$. Since the label of this edge is 
the first letter of a word in $[t]\varphi$ but not the first letter 
of any word in $[s]\varphi$, we have $[t]\varphi \not\subseteq [s]\varphi$.

Suppose now that $n > 1$ and that we have established the claim for 
$n - 1$. Now player I can select an edge originating in the 
root of $t$, labeled $a \in A$, with target the root of 
a subtree $t'$ such that for each $a$-labeled edge 
from the root of $s$ to the root of some subtree $s'$, 
player I wins the $(n-1)$-round game on $(t',s')$. By the 
induction hypothesis, this means that 
$[t']\varphi \not \subseteq [s']\varphi$
for all such subtrees $s'$.

Let $t_1,\ldots,t_k$ be up to isomorphism all those subtrees of $s$ 
whose roots are the targets of 
$a$-labeled edges originating in the root of $s$. We have  
$[t']\varphi \not \subseteq [t_i]\varphi$
for all $i$. Now $[t]\varphi$ contains $a(\#[t']\varphi \$)^k$ as a subset, and all the 
words in $[s]\varphi$ starting with $a$ are in the set 
$\{a\} \cup \bigcup_{i \in [k]}\bigcup_{m \geq 1} a(\#[t_i]\varphi \$)^m$.
For each $i \in [k]$, let $u_i$ be a word in $[t']\varphi$ which is not 
in $[t_i]\varphi$. Then the word $a\#u_1\$\ldots \#u_k\$$ is in $[t]\varphi$ 
but does not belong to $[s]\varphi$, since it does not belong to any
$a(\#[t_i]\varphi)^k$. Thus, $[t]\varphi \not \subseteq [s]\varphi$. \eop

\begin{prop}
\label{prop-emb2}
For each set $A$ there exist a ranked set $\Sigma$ and an injective dagger preserving 
(idempotent) grove theory morphism $\SRST_A \rightarrow \Reg_\Sigma$. 
\end{prop}

{\sl Proof.} 
Let $\Sigma_0 = A \cup \{\#,\$\}$, $\Sigma_2 = \{\sigma\}$, 
and let $\Sigma_n$ be empty if $n = 1$ or $n > 2$. For each $a\in A$,
consider a regular tree language $L_a: 1\rightarrow 1$ in $\Reg_\Sigma$ 
whose `frontier' is the context-free language described in the previous 
proof. (Such a regular tree language exists, since context-free 
languages are exactly the frontier languages of regular tree languages, 
see e.g. \cite{GecsegSteinby}.) For example, let 
$L_a = \{t_0 = a, 
t_1 = \sigma a \sigma \#  \sigma x_1 \$, 
t_2 = \sigma a \sigma \#  \sigma x_1 \sigma \$ \sigma \# \sigma x_1 \$,   \ldots \} $.
Then let $\psi: \SRST_A \rightarrow \Reg_\Sigma$ be the unique dagger preserving 
morphism of idempotent grove theories determined by the assignment
$a \mapsto L_a$ for all $a\in A$,
which exists by Corollary~\ref{cor-free}.
The morphism $\varphi$ 
constructed in the proof of Proposition~\ref{prop-emb1} factors through $\psi$ by the `frontier map'.
Since $\varphi$ is injective, so is $\psi$. 
 \eop

We are now ready to prove Theorem~\ref{thm-main}. 
Suppose first that the identity $t = t'$ holds in all power-set theories, 
where $t,t'$ are terms of sort $n \rightarrow p$. 
Since the theories $\Lang_A$ 
are particular power-set theories, $t = t'$ holds in all such theories as well as in 
all theories $\CF_A$, since $\CF_A$ embeds in $\Lang_A$. Any theory $\TreeLang_\Sigma$
embeds in a theory $\Lang_A$, and similarly, any theory $\Reg_\Sigma$ embeds in 
some theory $\CF_A$. Thus, if $t = t'$ holds in the theories $\Lang_A$ ($\CFL_A$, resp.),
then it holds in all theories $\TreeLang_\Sigma$ ($\Reg_\Sigma$, resp.). 
Suppose now that $t = t'$ holds in all theories $\CFL_A$. Then by Proposition~\ref{prop-emb2},
$t = t'$ holds in all theories $\SRST_A$, where $A$ is any set. 
This in turn implies by Proposition~\ref{prop-emb0} that 
$t = t'$ holds in all theories $\SRST_\Sigma$, where $\Sigma$ is a ranked set. 
Now, this yields that $t = t'$ holds in the theories $\SST_\Sigma$. Indeed, given $\Sigma$,
we can choose the ranked set $\Delta$ large enough such that for each $n$ there is 
surjective map $\Delta_n \rightarrow \SST_\Sigma(1,n)$. These maps determine a rank preserving 
function $\varphi :\Delta \rightarrow \SST_\Sigma$ which can be extended to a (necessarily surjective) dagger preserving morphism 
 $\varphi^\sharp: \SRST_\Delta \rightarrow \SST_\Sigma$
 of idempotent grove theories.  
Thus, any identity that holds in all theories $\SRST_\Sigma$ also 
holds in all theories 
$\SST_\Sigma$.
The same argument proves that if an 
identity holds in the theories $\SRST_\Sigma$ then it holds in 
all continuous idempotent grove theories. 
 The fact that if $t = t'$ holds in the theories $\SST_\Sigma$, then it 
also holds in all power-set theories now follows by noting that 
$\SRST_\Sigma$ embeds in $\SST_\Sigma$. \eop

\begin{remark} (Based on \cite{Essup}.)
The above proof also establishes the fact that an identity holds in all 
$\omega$-continuous idempotent grove theories iff it holds in all
theories $\SRST_\Sigma$ (or the theories mentioned in Theorem~\ref{thm-main}).

When $A$ is a poset with all countable suprema, the $\omega$-continuous functions 
over $A$ form an  $\omega$-continuous idempotent grove theory $\omega\Cont_A$.
A morphism $n \rightarrow p$ in this theory is an $\omega$-continuous function
$A^p \rightarrow A^n$ (note the reversal of the arrow), and composition is function 
composition (in the reverse order). For each $i \in [n]$, $n \geq 0$, the 
distinguished morphism $i_n: 1 \rightarrow n$ is the $i$th projection function 
$A^n \rightarrow A$. The constant $0_{n,p}$ is the constant function mapping all elements 
of $A^p$ to the least element of $A^n$, and $f + g$ is the pointwise supremum
of $f$ and $g$, for each $f,g: n \rightarrow p$ (i.e., $\omega$-continuous functions
$f,g: A^p \rightarrow A^n$). Note that the order $\leq$ becomes the pointwise partial order.
Since $\omega\Cont_A$ is a continuous idempotent grove theory, it comes with 
the least fixed point operation as dagger operation. 

Every $\omega$-continuous idempotent grove theory $T$ may be embedded in a theory
$\omega\Cont_A$. Given $T$, let $A = \prod_{p\geq 0}T(1,p)$, equipped with the 
pointwise partial order. The embedding maps a morphism $f: 1 \rightarrow n$ to the 
$\omega$-continuous function $A^p \rightarrow A$ defined by:
\begin{eqnarray*}
f((g_{1,p})_p,\ldots,(g_{n,p})_p)&=& 
(f \cdot \langle g_{1,p},\ldots,g_{n,p} \rangle)_p.
\end{eqnarray*}
By this embedding, we conclude that an identity holds in all continuous idempotent 
grove theories iff it holds in all theories of the sort $\omega\Cont_A$,
where $A$ is an $\omega$-continuous poset (or in fact a complete lattice, 
since every $\omega$-continuous poset may be embedded in a complete lattice).
\end{remark}

\section{Completeness}

We devote the rest of the paper to axiomatic issues. The results of this section 
are taken from the conference paper \cite{Essup} and are included here 
for two main reasons. First, \cite{Essup} uses a different formalism, the language 
of $\mu$-terms. Second, due to size limitations, \cite{Essup} does not contain
full proofs. The argument presented here is a modification of that
outlined in \cite{Essup}. 

\emph{Iteration theories}, introduced as a generalization of Elgot's iterative theories 
\cite{Elgot} and the rational and continuous theories \cite{ADJ,ADJrational} 
of the ADJ group,
provide an axiomatic framework to fixed point computations.
An iteration theory \cite{BEbook}
is a theory $T$ equipped with a dagger operation 
$^\dagger: T(n,n + p) \rightarrow T(n,p)$ for all $n,p \geq 0$, satisfying certain 
equational axioms such as the {\em fixed point equation}
\begin{eqnarray}
\label{eq-fp}
f \cdot \langle f^\dagger,\1{p}\rangle &=& f^\dagger
\end{eqnarray}
or the \emph{parameter equation} 
\begin{eqnarray}
\label{eq-parameter}
(f \cdot (\1{n} \os g))^\dagger &=& f^\dagger \cdot g,
\end{eqnarray}
where $f: n \rightarrow n + p$ and $g: p \rightarrow q$. 
For complete (equational) axiomatizations we refer to 
\cite{BEbook,Esgroup}. A morphism of iteration theories 
is a theory morphism preserving dagger. 
Iteration theory congruences are defined in the expected way.

A {\em grove iteration theory} \cite{BEbook,BET}
is an iteration theory which is a grove theory satisfying
\begin{eqnarray*}
\1{1}^\dagger &=& 0_{1,0}.
\end{eqnarray*} 
It is known that in a grove iteration theory, 
\begin{eqnarray*}
(\1{n} \os 0_p)^\dagger &=&  0_{n,p}
\end{eqnarray*} 
holds for all $n,p \geq 0$. 

It is possible to define a \emph{star operation} 
$^*:T(n,n+p) \rightarrow T(n,n+p)$ in any grove iteration theory.
When $f: n \rightarrow n + p$, we define
\begin{eqnarray*}
f^* &=& (f \cdot (\1{n} \os 0_n \os \1{p}) + (0_n \os \1{n} \os 0_p))^\dagger.
\end{eqnarray*}
It can be seen, cf. \cite{BEbook}, that in grove iteration theories $T$, 
$f^*$ is a solution of the equation 
\begin{eqnarray*}
x &=& f \cdot \langle x, 0_n \os \1{p} \rangle + (\1{n} \os 0_p),
\end{eqnarray*}
for all $f: n \rightarrow n + p$ in $T$.
When $p = 0$, this equation becomes $x = f\cdot x +\1{n}$.

A grove iteration theory is 
\emph{$\omega$-idempotent} if 
\begin{eqnarray*}
\1{1}^* &=& \1{1}
\end{eqnarray*} 
holds. 
Any $\omega$-idempotent grove iteration 
theory is an idempotent grove theory and thus
an idempotent grove iteration theory, Indeed, if $T$ 
is $\omega$-idempotent, then 
$$\1{1} + \1{1} = \1{1}\cdot \1{1}^* + \1{1} = \1{1}^* = \1{1}$$
and thus $f + f = (\1{1} +\1{1})\cdot f = \1{1} \cdot f = f$,
for all $f: 1 \rightarrow p$.  
Thus, we can define a partial order as above by $f \leq g$
iff $f + g = g$, for all $f,g: n \rightarrow p$. 
An idempotent grove iteration theory is \emph{ordered} if 
the dagger operation is monotone: 
\begin{eqnarray*}
f \leq g: n \rightarrow n + p &\Rightarrow & f^\dagger \leq g^\dagger: n \rightarrow p,
\end{eqnarray*}
or equivalently, if 
\begin{eqnarray*}
f &\leq (f + g)^\dagger,
\end{eqnarray*}
for all $f,g: n \rightarrow n + p$. 
It follows that composition is also monotone,
since in iteration theories, 
\begin{eqnarray*}
f \cdot g &=& (\1{n} \os 0_p)\cdot \langle 0_n \os f \os 0_q, 0_{n + p} \os g\rangle^\dagger,
\end{eqnarray*}
for all $f: n \rightarrow p$ and $g: p \rightarrow q$. (It can be seen that an 
idempotent grove theory is ordered iff composition and the scalar dagger 
operation $f \mapsto f^\dagger$, $f : 1 \rightarrow 1 +p$ are monotone.)

A morphism of grove iteration
theories is both a grove theory morphism and an iteration theory
morphisms. A congruence of grove iteration theories is a grove
theory congruence which is an iteration theory congruence.
Morphisms and congruences of $\omega$-idempotent 
grove iteration theories are defined in the same way. 

We have already defined the congruence $\equiv_s$ of simulation
equivalence on $\ST_\Sigma$. A closely related congruence is 
\emph{bisimulation equivalence}, or {\em bisimilarity}. 
Recall \cite{Milner,Park} that a simulation $R: f \rightarrow g$ 
for trees $f,g\in \ST_\Sigma$ is a bisimulation if the
relational inverse of $R$ is a simulation $g \rightarrow f$. 
We say that trees $f,g: 1 \rightarrow p$ are bisimilar if there is
a bisimulation $f \rightarrow g$. When $f = (f_1,\ldots,f_n): n \rightarrow p$
and $g = (g_1,\ldots,g_n): n \rightarrow p$, then we say that $f$ 
is bisimilar to $g$ is each $f_i$ is bisimilar to $g_i$.
As mentioned above, bisimilarity, denoted $\equiv_b$,
is a grove iteration theory congruence \cite{BEbook}, moreover, $\BST_\Sigma =
\ST_\Sigma/\equiv_b$ is an $\omega$-idempotent grove iteration 
theory as is $\BRST_\Sigma = \RST_\Sigma /  \equiv_b$.
The bisimilarity class of a tree $f: n \rightarrow p$ in $\ST_\Sigma$
(or in $\RST_\Sigma$) will be denoted $[f]_b$. Note that we 
may embed $\Sigma$ into $\BRST_\Sigma$ or $\BST_\Sigma$ by the function 
mapping a letter $\sigma \in \Sigma_n$ into the bisimilarity 
class of $\sigma$ seen as a tree $1 \rightarrow n$.

The following result was proved in \cite{BET,BEbook} (at least in the case 
when each letter in $\Sigma$ has rank 1).

\begin{thm}
\label{thm-free-bisim}
For each ranked alphabet $\Sigma$,  $\BRST_\Sigma$  is the 
free $\omega$-idempotent  grove iteration theory, freely generated by $\Sigma$.
In more detail, given any $\omega$-idempotent iteration theory $T$
and rank preserving function $\Sigma\rightarrow T$, there is a unique morphism
of $\omega$-idempotent grove iteration theories $\varphi^\sharp: \BRST_\Sigma \rightarrow T$ 
extending $\varphi$.  
\end{thm}

We will use Theorem~\ref{thm-free-bisim} in our characterization 
of $\SRST_\Sigma$ as the free ordered $\omega$-idempotent grove theory,
freely generated by $\Sigma$. But first we need some preparations. 
Recall that a morphism $\tau: f \rightarrow g$ for $f,g: 1 \rightarrow p$ in $\ST_\Sigma$ 
is a function $V_f \rightarrow V_g$ with appropriate properties so that we may
extend $\tau$ to a function $E_f \rightarrow E_g$. The definition of the extension 
is unique for edges $v \rightarrow (v_1,\ldots,v_n)$ with $n \geq 1$,
but there might be several options to define the image of an edge 
with empty target sequence. Call $\tau$ 
\emph{locally surjective} if has an extension to edges 
such that for every vertex $v$ of $f$, the edges of $f$ with source
$v$ are surjectively mapped onto the edges of $g$ with source
$v\tau$. Clearly, every locally surjective morphism is surjective. 
Moreover, we have: 

\begin{lem}
Suppose that $\Sigma$ is a ranked alphabet and $f,g: 1 \rightarrow p$ 
in $\ST_\Sigma$. Then a morphism $\tau: f \rightarrow g$ is a bisimulation iff 
$\tau$ is locally surjective.
\end{lem}

\begin{lem}
\label{lem-embed}
Suppose that $\Sigma$ is a ranked alphabet and 
$f,g : 1 \rightarrow p$ in $\RST_\Sigma$ with $f \sqsubseteq_s g$.
Then there is a regular tree $f': 1 \rightarrow p$ in $\RST_\Sigma$
 such that $f'\equiv_b g$ and 
$f$ embeds in $f'$ by an injective synchronization tree morphism.
\end{lem}

{\sl Proof.} First note that since $f \sqsubseteq_s g$,
there is a morphism $\tau: f \rightarrow g$ such that if $f|_u$ is isomorphic to $f|_v$ 
and $g|_{u\tau}$ is isomorphic to $g|_{v\tau}$ for some vertices $u,v$ of $f$, 
then $f|_u\tau$  and $f|_v\tau$ are also isomorphic. Thus, we can write each subtree 
$g|_{u\tau}$ in the form $f|_u\tau + g_u$ for a (unique) tree $g_u$. 
Let $f'$ be obtained from $f$ by `adding' $g_u$ as a summand to $f|_u$,
for each vertex $u$ of $f$, so that $f'|_u = f|_u + g_u$ for each vertex $u$
of $f$. Then clearly $f$ embeds in $f'$ and $f' \equiv_b f$, since there is a 
locally surjective morphism $f' \rightarrow g$.
\eop   

\begin{lem}
\label{lem-embed2}
Suppose that $\approx$ is a grove iteration theory congruence
of $\BRST_\Sigma$ such that the quotient $\BRST_\Sigma/\approx$ 
is an ordered grove iteration theory morphism. For all $f,g: n\rightarrow p$ in 
$\RST_\Sigma$, define $f \preceq g$ when $[f]_b + [g]_b \approx [g]_b$.
Moreover, define $f \approx g$ iff $[f]_b \approx [g]_b$, so that 
$f \approx g$ iff $f \preceq g$ and $g \preceq f$.

Let $f,g: 1 \rightarrow p$ in $\RST_\Sigma$. 
If $f:1 \rightarrow p$ embeds in $g: 1 \rightarrow p$ by an injective morphism, then $f \preceq g$.
\end{lem}

{\sl Proof.} Without loss of generality, we may choose an embedding $\tau: f \rightarrow g$
such that whenever $f|_u$ is isomorphic to $ f|_v$ and $g|_{u\tau}$
is isomorphic to $g|_{v\tau}$, for some
vertices $u,v$ of $f$, then $f|_u\tau$ and $f|_v\tau$
are also isomorphic.
Let $f_1=f,\ldots,f_n$ and $g_1=g,\ldots,g_m$ be, up to isomorphism, all the subtrees 
of $f$ and $g$, respectively. Let us label a vertex $u$ of $f$ by the pair $(i,j)$
if $f|_u = f_i$ and $g|_u = g_j$. Let $K$ denote the set of all pairs 
that label some vertex $u$, and consider a bijection $\psi$  between 
the sets $K$ and 
$[k]$, where $k$ denotes the number of elements of $K$. 
We may assume that $(1,1)\psi = 1$. Let $\psi'$ denote the inverse of $\psi$.
Now using $f$, let us construct a tree $a: k \rightarrow k +p $, $a = \langle a_1,\ldots,a_k\rangle$.
Consider a vertex $u$ of $f$ labeled $(i,j)$, say, together with all edges $e$ 
originating in $u$. For each $\sigma \in \Sigma_r$ that occurs as the label 
of an edge, and for each sequence $(i_1,j_1),\ldots,(i_r,j_r)$ of elements of $K$,
let $\ell = \ell(\sigma,(i_1,j_1),\ldots,(i_r,j_r))$ denote the number of 
outgoing edges of $u$
labeled $\sigma$ with target vertices labeled $(i_1,j_1),\ldots,(i_r,j_r)$.
Consider the tree $\ell\cdot \sigma\cdot \rho$, where $\rho$ is the 
base morphism $r \rightarrow k$ corresponding to the function $[r]\rightarrow [k]$ with
$s \mapsto (i_s,j_s)\psi$, for all $s \in [r]$. Moreover, for each $j\in [p]$,
let $\ell_j$ denote the number of edges originating in $u$ labeled $\ex_j$.
Then the component $a_{(i,j)\psi}$ of $a$ corresponding to 
$(i,j)\psi$ is the sum of all such trees $(\ell \cdot \sigma \cdot \rho)\os 0_p$ 
and $0_k \os (\ell_j \cdot j_p)$. 

Having constructed $a: k \rightarrow k + p$, we construct $b: k \rightarrow k + p$ in the same way 
using the tree $g$. Clearly, $a \preceq b$, since $b = a + c$ 
for some $c$. For each $s \in [k]$, let $i_s$ be the first component 
of $s\psi'$, and $j_s$ the second. Then the fixed point equation $x = a \cdot \langle x,\1{p}\rangle$ has $\langle f_{i_1},\ldots,f_{i_k}\rangle$ as its unique solution.
Similarly, the equation $y = b \cdot \langle y,\1{p}\rangle$ has 
$\langle g_{j_1},\ldots,g_{j_k}\rangle$ as its unique solution.
In particular, $f = 1_k \cdot a^\dagger$ and $g = 1_k \cdot b^\dagger$. 
Since $a \preceq b$, we conclude that $f \preceq g$.
\eop

\begin{thm}
\label{thm-free simulation}
For each ranked set $\Sigma$, 
$\SRST_\Sigma$ is the free ordered $\omega$-idempotent grove iteration theory,
freely generated by $\Sigma$. 
\end{thm}

{\sl Proof.} 
Let $\approx$ denote the least grove iteration theory congruence
of $\BRST_\Sigma$ such that the quotient $\BRST_\Sigma/\approx$ 
is an \emph{ordered} grove iteration theory. For all $f,g: \rightarrow p$ in 
$\RST_\Sigma$, define $f \preceq g$ when $[f]_b + [g]_b \approx [g]_s$.
Moreover, define $f \approx g$ iff $[f]_b \approx [g]_b$, so that 
$f \approx g$ iff $f \preceq g$ and $g \preceq f$.

Our goal is to show that the relations $\sqsubseteq_s$ and $\preceq$ on $\RST_\Sigma$
are equal.
It is clear that $\sqsubseteq_s$ is included in $\preceq$. To complete 
the proof, it suffices to show that $f \preceq g$ whenever $f,g:1 \rightarrow p$
in $\RST_\Sigma$ with $f \sqsubseteq_s g$.  But if $f \sqsubseteq_s g$,
then by Lemma~\ref{lem-embed}
there exists $f': 1 \rightarrow p$ in $\RST_\Sigma$ with $f' \equiv_b g$ and such that there is 
an injective synchronization tree morphism $f \rightarrow f'$.  
But then $f' \approx g$  by Theorem~\ref{thm-free-bisim}, 
and $f \preceq f'$ by Lemma~\ref{lem-embed2}. It follows that 
$f \preceq g$. \eop

A \emph{idempotent Park grove theory} is an ordered idempotent grove theory
equipped with a dagger operation satisfying the fixed point identity (\ref{eq-fp}),
the parameter identity (\ref{eq-parameter}) and the \emph{fixed point induction rule}:
\begin{eqnarray*}
f \cdot \langle g ,\1{p}\rangle \leq g &\Rightarrow& f^\dagger \leq g,
\end{eqnarray*}
for all $f: n \rightarrow n + p$ and $g: n \rightarrow p$. For example, all $\omega$-continuous 
idempotent grove theories are idempotent Park grove theories. It is known,
cf. \cite{Espark}, that every idempotent Park grove theory is an 
ordered $\omega$-idempotent iteration theory. 
A morphism of idempotent Park grove theories is an idempotent grove theory 
morphism which preserves dagger. 

From Theorem~\ref{thm-free simulation} we have: 

\begin{cor}
\label{cor-free simulation}
For each ranked set $\Sigma$, 
$\SRST_\Sigma$ is the free idempotent Park grove theory,
freely generated by $\Sigma$. 
\end{cor}

\begin{cor}
The following are equivalent for an identity $t= t'$ between terms $t,t': n \rightarrow p$
involving the theory operations, dagger, and sum.
\begin{itemize}
\item $t=t'$ holds in all $\omega$-continuous idempotent grove theories.
\item $t=t'$ holds in all ordered $\omega$-idempotent grove iteration theories.
\item $t = t'$ holds in all idempotent Park grove theories.
\end{itemize}
\end{cor}

\begin{remark}
Adding the equations 
\begin{eqnarray*}
f \cdot (g + h) &=& f \cdot g + f \cdot h\\
f \cdot 0_{p,q} &=& 0_{n,p},
\end{eqnarray*}
for all $f:  n \rightarrow p$ and $g,h: p \rightarrow q$, the resulting system is 
known to be complete 
for (matrix) theories over $\omega$-continuous idempotent semirings, or 
regular languages, or theories of $\omega$-continuous functions over 
$\omega$-continuous semirings or complete lattices  
preserving binary (or countable)  suprema, or theories of binary relations. 
See \cite{BEbook}, where original references may be found. 
\end{remark}


\end{document}